\documentclass[aps,twocolumn,10pt,showpacs]{revtex4}

\usepackage{amsmath}
\usepackage{amssymb}
\usepackage{graphicx}
\usepackage{dcolumn}

\begin{document}

\title{Thermodynamics of parametric dark energy models }
\author{Samantha Rath} 
\thanks{samantha.rath12@gmail.com}
\affiliation
{Department of Physics and Astrophysics, University of Delhi,
Delhi-110007, India. }

\date{\today}

\begin{abstract}
A comparative study of a set of parametric dark energy models is performed
by studying the evolution of dark energy both in the past and future epochs.
In addition, the age of the universe and time till the distant future ($%
a=1000$) are estimated. The validity of generalized second law of
thermodynamic in different parametric models is also ascertained.
\end{abstract}
\maketitle

\section{Introduction}

In contemporary cosmology, the story of dark energy is quite checkered and
enigmatic in nature. The fact that dark energy is a dominant component of
cosmic constituents has been well established directly from the observation
of acceleration in cosmic expansion as indicated by SN Ia data \cite%
{perl98,ries04} and indirectly through different cosmological probes, namely
CMB anisotropy measurements \cite{sper03,scra03}, BAO and \cite%
{eise05,perc07} strong \cite{suyu10} as well as weak \cite{schr10}
gravitational lensing. As the acceleration of expansion due to repulsive
dark energy can be provided by the cosmological constant $\Lambda $, the
standard cold dark matter (SCDM) model has consequently been replaced by $%
\Lambda $CDM model. In spite of the fact that the $\Lambda $CDM model
provides the best fit to all the available cosmological data, two well known
theoretical issues, namely coincidence and fine tuning remain to be
explained \cite{wein13}.

In the absence of definitive answer regarding the source of dark energy,
several alternative scenarios based on scalar field, string theory, modified
gravity, Chaplygin gas, polytropic gas and interacting dark energy have been
explored \cite{cope06,faro10,noji11,capo11,bamb12,bamb15,cai16,noji17,hute18}%
. Within scalar field models, quintessence, phantom, k-essence, DBI-essence,
H-essence, tachyon, dilaton, quintom and ghost condensate have been
proposed. An exhaustive list of parametric models has been reported by Pacif 
\cite{paci20}. In case of parametric models with varying $w=w(z)$, different
parametrizations of $w(z)$, namely one index \cite{gong05}, two index such
as, linear parametrization \cite{coor99,asti01,well02,upad05},
Chevallier-Polarski-Linder (CPL) \cite{chev01,lind03}, \
Jassal-Bagla-Padmanabhan (JBP) \cite{jass04}, generalized CPL as well as JBP 
\cite{liu08}, Feng-Shen-Li-Li (FSLL) \cite{feng12}, Barboza-Alcaniz (BA) 
\cite{barb12}, square-root \cite{pant16}, Ma-Zhang (MZ) \cite{ma11},
logarithmic \cite{efst99,wett04,ma11,feng11,sell13} and oscillating \cite%
{lazk10,pan18}, three index \cite{lind05,alam04} and four index \cite%
{hann04,lee05} parametrizations have been proposed.

Usually, the parameters of a specific model are marginalized in the red
shift range $z=0-1090$ employing available cosmological data namely SN Ia,
BAO, CMB, $H(z)$ data. By adopting this procedure, it turns out that the $%
\Lambda $CDM model and parametric models exhibit similar dynamics of dark
energy. The age of the universe is an integrated quantity. Hence, it can be
used in the comparative study of different parametric models. As the fitting
procedure cannot include future data, it will also be interesting to explore
the future behavior of these parametric models. In Ref. \cite{card17}, two
requirements of generalized second law (GSL) of thermodynamics as
thermodynamic viability of dark energy models are given by $S^{\prime
}(a)\geqslant 0$ and in the far future $S^{\prime \prime }(a)\leq 0$. We
calculate first derivative $S^{\prime }(a)$ and second derivative $S^{\prime
\prime }(a)$ of entropy $S$ and thereby investigate the thermodynamic
viability of parametric dark energy models.

\section{Thermodynamics of dark energy models}

In a spatially flat ($k=0$) universe, the constituents of cosmic energy
density are non-relativistic matter (baryonic and dark matter), neutrinos,
CMB photons, and dark energy. In this work, contribution of neutrinos to the
cosmic energy density is neglected. With the equation of state $P=\omega
\rho $, the Friedmann-Raychaudhury equations are written as 
\begin{eqnarray}
H^{2}(a)\;\; &=&\frac{8\pi G}{3}\sum_{i}\rho _{i}(a) \\
\frac{\overset{..}{a}}{a} &=&-\frac{4\pi G}{3}\;\sum_{i}\rho _{i}(a)\left[
1+3w_{i}(a)\right]
\end{eqnarray}%
The equation of state parameters $\omega _{M}=0$ and $\omega _{R}=1/3$ for
matter and radiation, respectively. In Section 3, the dark energy parameter $%
\omega _{X}$ of different parametric models is discussed.

The energy densities $\rho _{i}(a)$ and density parameters $\Omega _{i}(a)$
are related by 
\begin{eqnarray}
\rho _{i}(a) &=&\rho _{0i}f_{i}(a) \\
\Omega _{i}(a) &=&\frac{\Omega _{0i}f_{i}(a)}{E^{2}}
\end{eqnarray}%
where the dark energy function $f_{i}(a)=$ $\rho _{i}(a)/\rho _{0i}$ and $%
E(a)$ are defined by%
\begin{eqnarray}
f_{i}(a)\;\; &=&\;\;\exp \left[ -3\int_{1}^{a}{\frac{1+w_{i}(a^{\prime })}{%
a^{\prime }}da^{\prime }}\right] \\
E^{2}(a) &=&\frac{H^{2}(a)}{H_{0}^{2}}=\sum_{i}\Omega _{i0}f_{i}(a)
\label{e2}
\end{eqnarray}%
Using Eq. (\ref{e2}), the second Friedmann-Raychaudhury equation can be
written as 
\begin{equation}
\frac{2q-1}{3}=\frac{\Omega _{R}}{3}+w_{X}\Omega _{x}  \label{qt}
\end{equation}%
The age of the universe is calculated by using 
\begin{widetext}
\begin{eqnarray}
t_{0}\;\; &=&\;\;\int_{0}^{1}\frac{da}{aH(a)}  \notag \\
&=&H_{0}^{-1}\;\int_{0}^{1}\frac{da}{a\left[ \Omega _{0M}a^{-3}+\Omega
_{0R}a^{-4}+\Omega _{0X}f_{x}(a)\right] ^{1/2}}  \label{age}
\end{eqnarray}%
\end{widetext}
\ \ \ \ \ \ \ \ \ 

According to the GSL of thermodynamics, two requirements of thermodynamic
viability of dark energy models are given by $S^{\prime }(a)\geqslant 0$ and
in the far future $S^{\prime \prime }(a)\leq 0$. The total entropy of the
universe $S(a)$ is due to the entropy of horizon $S_{H}(a)$ as well as
entropy of cosmic fluids within the horizon, namely matter $S_{M}(a)$,
radiation $S_{R}(a)$ and dark energy $S_{X}(a)$. Explicitly, the entropy $%
S_{H}(a)$, $S_{M}(a)$, $S_{R}(a)$ and $S_{X}(a)$ are written as follows \cite%
{card17}.%
\begin{eqnarray}
S_{H}(a) &=&\frac{k}{4}\frac{A_{H}}{l_{p}^{2}}  \notag \\
&=&\frac{\pi kc^{2}}{4H^{2}} \\
S_{M}(a) &=&knV_{H}  \notag \\
&=&\frac{4\pi kc^{3}n_{0}}{3a^{3}H^{3}}
\end{eqnarray}%
Entropy of radiation and dark energy are calculated using%
\begin{equation}
TdS_{i}=d(\rho _{i}V_{H})+\rho _{i}\omega _{i}d(V_{H})
\end{equation}

In dimensionless form 
\begin{equation}
\widetilde{S}_{\alpha }(a)=\left( \frac{4G\hbar H_{0}^{2}}{3kc^{5}}\right)
S_{\alpha }
\end{equation}%
where $\alpha =H$, $M,R$ and $X$ denote horizon, matter, radiation, and dark
energy,respectively, the first derivative of total entropy of the universe $%
\widetilde{S}^{\prime }(a)$ is written as%
\begin{equation}
\widetilde{S}^{\prime }(a)=\widetilde{S}_{H}^{\prime }(a)+\widetilde{S}%
_{M}^{\prime }(a)+\widetilde{S}_{R}^{\prime }(a)+\widetilde{S}_{X}^{\prime
}(a)
\end{equation}%
with 
\begin{widetext}
\begin{eqnarray}
\widetilde{S}_{H}^{\prime }(a) &=&\frac{4\pi }{aE^{2}(a)}\left[
1+\sum_{i}w_{i}(a)\Omega _{i}(a)\right]  \label{ssh} \\
\widetilde{S}_{M}^{\prime }(a) &=&\left( \frac{H_{0}\hbar }{kT_{0M}}\right) 
\frac{1}{a^{4}E^{3}(a)}\left[ 1+3\sum_{i}w_{i}(a)\Omega _{i}(a)\right]
\label{ssm} \\
\widetilde{S}_{i}^{\prime }(a) &=&\left( \frac{H_{0}\hbar }{kT_{i}}\right) 
\frac{\Omega _{i}(a)\left( 1+w_{i}(a)\right) }{aE(a)}\left[
1+3\sum_{j}w_{j}(a)\Omega _{j}(a)\right]  \label{ssrx}
\end{eqnarray}%
\end{widetext}
Here, ``$i$" denotes radiation/dark energy. Temperature of cosmic fluids are
defined as%
\begin{eqnarray}
T_{0M} &=&\left( \frac{3c^{2}H_{0}^{2}}{8\pi Gkn_{0}}\right) \\
T_{i} &=&T_{0i}a^{3}f_{i}(z)
\end{eqnarray}%
where $n_{0}\thicksim 10^{-6}$ m$^{-3}$, $T_{0R}=2.725$ K and%
\begin{equation}
T_{0X}=\tau _{X}\frac{H_{0}\hbar }{k}
\end{equation}%
with an arbitrary parameter $\tau _{X}$. Finally, the second derivative of
total entropy $\widetilde{S}^{\prime \prime }(a)$ can be calculated from $%
\widetilde{S}^{\prime }(a)$ in a straight forward manner.

\section{Parametric models}

In addition to $\Lambda $CDM, quintessence and phantom models, a set of two
index parametric models in which $w(z)$ is bounded in both distant past ($%
a\rightarrow 0$) and far future ($a\rightarrow \infty $) are considered.
Although, the choice of parametrizations is arbitrary, they are
representative of different scenarios. With $w=$ constant, there are three
possible scenarios, namely $\Lambda $CDM, quintessence and phantom models.
In general, the dark energy function $f_{X}(a)$ is given by 
\begin{equation}
f_{X}(a)=a^{-3(1+w)}  \label{fzcon}
\end{equation}

\vskip 5mm
\textbf{I. }$\mathbf{\Lambda }$\textbf{CDM} \textbf{\ model}

In $\Lambda $CDM model, $w(a)=-1$ and dark energy function is given by 
\begin{equation}
f(a)=1
\end{equation}%
with $\Omega _{0M}=0.30$ \cite{blan22}.

\vskip 5mm
\textbf{II. Quintessence model}

In quintessence models, scalar fields with constant $w>-1$ are considered.
The dark energy function $f_{X}(a)$ is same as Eq. (\ref{fzcon}) and $%
w=-0.80 $ with $\Omega _{0M}=0.30$.

\vskip 5mm
\textbf{III. Phantom model}

Scalar fields with $w<-1$ are called phantom fields and in general, $f_{X}(a)
$ given by 
\begin{equation}
f_{X}(a)=a^{-3(1+w)}
\end{equation}%
In this work, $w=-1.2$ with $\Omega _{0M}=0.30$ is employed.

\vskip 5mm
\textbf{IV. Upadhye-Ishak-Steinhardt (UIS) parametrization}

In Ref. \cite{upad05}, Upadhye \textit{et al}. have parametrized $w(a)$ as 
\begin{equation}
w(a)=\left\{ 
\begin{array}{lll}
w_{0}+w_{1}(1-a)/a & if & a>0.5 \\ 
w_{0}+w_{1} & if & a\leq 0.5%
\end{array}%
\right.
\end{equation}%
and the corresponding the dark energy evolution function $f_{x}(a)$ is given
by 
\begin{widetext}
\begin{eqnarray}
f_{X}(a) &=&a^{-3(1+w_{0}-w_{1})}\exp \left[ \frac{3w_{1}(1-a)}{a}\right]
\qquad if\text{ }a>0.5  \notag \\
&=&a^{-3(1+w_{0}+w_{1})}\exp \left[ 3w_{1}(1-2\ln 2)\right] \qquad if\text{ }%
a\leq 0.5
\end{eqnarray}%
\end{widetext}
By $\chi ^{2}$ minimization of SN Ia `gold set', galaxy power spectrum and
CMB power spectrum data, the $1\sigma $ and $2\sigma $ constraints on dark
energy parameters are $w_{0}=-1.38_{-0.20-0.55}^{+0.08+0.30}$, $%
w_{1}=1.2_{-0.16-1.06}^{+0.40+0.64}$ and $\Omega _{0M}=0.31$. With these
parameters, the dark energy has phantom origin.

\vskip 5mm
\textbf{V. Feng-Shen-Li-Li (FSLL(I)) parametrization}

Using the parametrization given by Feng \textit{et al}. \cite{feng12} 
\begin{equation}
w(a)=w_{0}+\frac{w_{1}a(1-a)}{1-2a+2a^{2}}
\end{equation}%
the dark energy function $f_{X}(a)$ is written as 
\begin{widetext}
\begin{equation}
f_{X}(a)=a^{-3(1+w_{0})}(1-2a+2a^{2})^{3w_{1}/4}\exp \left[ \frac{3w_{1}}{2}%
\tan ^{-1}(\frac{1-a}{a})\right] \qquad
\end{equation}%
\end{widetext}
Global fitting of data on SNIa with 557 samples, WMAP7 data, SDSS DR7 and
perturbation of dark energy, the $1\sigma $ and $2\sigma $ constraints on
dark energy parameters are $w_{0}=-1.0148_{-0.5907-0.7247}^{+0.5183+0.6482}$%
, $w_{1}=-0.0155_{-2.0833-2.7378}^{+2.1257+2.6973}$, $\Omega
_{DM}h^{2}=0.1094_{-0.0133-0.0163}^{+0.0126+0.0155}$ and $\Omega
_{b}h^{2}=0.0223_{-0.0014-0.0019}^{+0.0016+0.0021}$.

\vskip 5mm
\textbf{VI. Feng-Shen-Li-Li (FSLL(II)) parametrization}

Second parametrization Feng \textit{et al}. \cite{feng12} given by 
\begin{equation}
w(a)=w_{0}+\frac{w_{1}(1-a)^{2}}{1-2a+2a^{2}}
\end{equation}%
results in the dimensionless dark energy function $f_{X}(a)$ given by
\begin{widetext} 
\begin{equation}
f_{X}(a)=a^{-3(1+w_{0}+w_{1})}(1-2a+2a^{2})^{3w_{1}/4}\exp \left[ -\frac{%
3w_{1}}{2}\tan ^{-1}(\frac{1-a}{a})\right] \qquad
\end{equation}%
\end{widetext}
By the same procedure as FSLL (I), the $1\sigma $ and $2\sigma $ constraints
on dark energy parameters are $%
w_{0}=-1.0214_{-0.2846-0.3373}^{+0.3093+0.4287}$, $%
w_{1}=-0.0113_{-2.5469-3.9020}^{+0.9683+1.1182}$, $\Omega
_{DM}h^{2}=0.1107_{-0.0141-0.0172}^{+0.0118+0.0161}$ and $\Omega
_{b}h^{2}=0.0225_{-0.0017-0.0021}^{+0.0015+0.0018}$.

\vskip 5mm
\textbf{VII. Barboza-Alcaniz (BA) parametrization}

Barboza and Alcaniz have parametrized $w(a)$ \cite{barb12} as 
\begin{equation}
w(a)=w_{0}+w_{1}\frac{(1-a)}{1-2a+2a^{2}}
\end{equation}%
and $f_{X}(a)$ is given by 
\begin{equation}
f_{X}(a)=a^{-3(1+w_{0}+w_{1})}(1-2a+2a^{2})^{3w_{1}/2}
\end{equation}%
The $1\sigma $ (best fit) values of parameters for fitting SN Ia Hubble
diagram by JLA sample, BAO (6dFGS), SDSS DR7, CMB (Plank13) and $H(a)$ data
are $w_{0}=-0.947_{-0.453}^{+0.339}$ ($-1.010$), $%
w_{1}=-0.013_{-0.149}^{+0.143}$ ($-0.076$), $\Omega
_{0M}=0.256_{-0.087}^{+0.058}$ ($0.290$) and $h=0.707_{-0.060}^{+0.097}$ ($%
0.699$).

\vskip 5mm
\textbf{VIII. Ma-Zhang (MZ) parametrization}

Ma and Zhang have parametrized $w(a)$ \cite{ma11} as 
\begin{equation}
w(a)=w_{0}+w_{1}\left[ a\ln \left( \frac{1+a}{a}\right) -\ln 2\right] 
\end{equation}%
and the resulting dark energy evolution function $f_{X}(a)$ is written as 
\begin{widetext}
\begin{equation}
f_{X}(a)=a^{-3(1+w_{0}-w_{1}\ln 2)}2^{6w_{1}}(1+a)^{-3w_{1}}\left( \frac{1+a%
}{a}\right) ^{-3w_{1}a}
\end{equation}%
\end{widetext}
With $\chi ^{2}$ minimization, the marginalized parameters ($95.4\%$ CL)
using SNIa (557 Union2), WMAP, and BAO (SDSS DR7) are $%
w_{0}=-1.067_{-0.155}^{+0.234}$, $w_{1}=-1.049_{-0.896}^{+5.706}$ and $%
\Omega _{M}^{0}=0.280_{-0.028}^{+0.032}$ and $h=0.697_{-0.026}^{+0.031}$.

\vskip 5mm
\textbf{IX. Pan-Saridakis-Yang (PSY1) parametrization}

Using the second parametrization of $w(a)$ given by Pan \textit{et al.} \cite%
{pan18}, 
\begin{equation}
w(a)=w_{0}+b\left[ 1-\cos (\ln a)\right] 
\end{equation}%
$f_{X}(a)$ is written as 
\begin{equation}
f_{X}(a)=a^{-3(1+w_{0}+b)}\exp \left[ 3b\sin (\ln a)\right] 
\end{equation}%
Using the same procedure as in PSY1 parametrizations, the marginalized dark
energy parameters with mean $\pm 1\sigma \pm 2\sigma \pm 3\sigma $ (best fit
values) are $w_{0}=-1.0078_{-0.032-0.059-0.080}^{+0.023+0.068+0.094}$ ($%
-1.0031$), $b=-0.1468_{-0.142-0.555-0.803}^{+0.275+0.431+0.511}$ ($-0.1127$%
), $\Omega _{0M}=0.306_{-0.009-0.017-0.020}^{+0.008+0.017+0.025}$ ($0.308$),
and $H_{0}=68.05_{-0.90-2.02-2.68}^{+1.20+1.77+2.25}$ (67.84).

\vskip 5mm
\textbf{X. Pan-Saridakis-Yang (PSY2) parametrization}

Using the parametrized $w(a)$ given by Pan \textit{et al.} \cite{pan18} 
\begin{equation}
w(a)=w_{0}-b\sin \left( \ln a\right)
\end{equation}%
the dark energy function $f_{X}(a)$ is found to be 
\begin{equation}
f_{X}(a)=a^{-3(1+w_{0})}\exp \left[ 3b(1-\cos (\ln a)\right]
\end{equation}%
The dark energy parameters have been marginalized by using SNIa (JSL
sample), BAO, CMB, red shift space distortions, weak gravitational lensing,
and Hubble parameters. Constrained parameters with mean $\pm 1\sigma \pm
2\sigma \pm 3\sigma $ (best fit values) are $%
w_{0}=-0.9817_{-0.0616-0.1032-0.1390}^{+0.0535+0.0938+0.1175}$ ($-1.0444$), $%
b=-0.0114_{-0.0319-0.0809-0.1071}^{+0.0378+0.0739+0.1001}$ ($-0.0144$), $%
\Omega _{0M}=0.311_{-0.010-0.019-0.024}^{+0.011+0.019+0.023}$ ($0.300$) and $%
H_{0}=67.32_{-1.39-1.95-2.37}^{+1.09+2.22+2.86}$ ($68.74$).

\section{Results and Discussions}

In order to ascertain the similarities and differences between $\Lambda $CDM
and various parametric models, we compare the evolution of $\Omega _{X}$ due
to different parametrizations both in the past and future epochs. In the
past epoch, the dark energy density parameter $\Omega _{X}$ is calculated at
red-shifts of a few astrophysical epochs, namely $1\ \leq\ z\ \leq \ 3$ (galaxy
formation era ), $z=1090$ (last scattering surface), and $z\;=\;10^{10}$ (BBN). 
Constraints on dark energy density parameters during galaxy formation era, 
LSS and BBN are $\Omega _{X}\;<\;0.5$ \cite{free03}, $(\Omega _{X})_{dec}\;<\;0.1$ \cite%
{upad05,cald04} and $(\Omega _{X})_{BBN}\;<\;0.14$ \cite{johr02} ($(\Omega
_{X})_{BBN}\;<\;0.21$ \cite{cybu05}), respectively. These constraints have
been obtained such that during the galaxy formation era, $\Omega _{X}$ $<$ $%
\Omega _{M}$ and the observed He abundance in the universe should not be
disturbed due to the presence of dark energy until BBN epoch. The evolution
of $\Omega _{X}$ in the future epoch is reported at a few arbitrary
redshifts corresponding to $a=,0.25,0.50$ and $1000$.

Parameters of the above mentioned models are fitted to different sets of the then
available cosmological data. In most of the $\chi ^{2}$ minimization but for
Ref. \cite{card17}, contribution of radiation has not been taken into
account. The inclusion of radiation may play significant role while fitting data 
of early universe, such as BBN.
In the present analysis, we include the contribution of radiation
with $\Omega _{0R}=2.475\times 10^{-5}/h^{2}$. Although, the Hubble
constant has been fitted with a few parametrizations, the results of
the present work are almost independent of the magnitude of $h$ and hence, $%
h=0.699$ \cite{card17} is used for convenience. 

In Table 1, transition redshift $z_{T}$, dark energy density parameter at
transition ($\Omega _{X}$)$_{T}$, $\Omega _{X}$ ($a=0.5-0.25$) during galaxy
formation era, and $\Omega _{X}$ ($a=0.0009$) at LSS ($z=1090$) are
displayed. Also, the age of the universe (in dimensionless units), 
$tH_{0}$ is given in the last
column of the same Table. It is noticed that in comparison to $\Lambda $CDM
model, the transition redshift $z_{T}$ in all models but for UIS and BA
differ by less than 5\%. In fact, UIS parameters have been marginalized with
earlier cosmological data set \cite{upad05}. However, variation in ($\Omega
_{X}$)$_{T}$ due to different parametrizations is within 5\% except for
quintessence, phantom, and UIS parametrizations. During galaxy formation
epoch, the values of $\Omega _{X}$ show the same trend albeit the constraint
on $\Omega _{X}$ $<0.5$ corresponding to $a=0.5--0.25$ is satisfied by all the parametric
models. Further, all the parametric models are consistent with LSS and BBN
constraints as $\Omega _{X}\rightarrow 0$.

\begin{table}[htbp]
\caption{Evolution of dark energy parameter $\Omega _{X}$ in the past
epoch.}
\begin{tabular}{llllllllcllll}
\hline\hline
{\small Models} &  &  & ${\small z}_{T}$ &  &  & {\small (}$\Omega _{X}$%
{\small )}$_{T}$   &  & \multicolumn{3}{c}{${\small \Omega }_{X}$} &  &   $%
{\small t}_{0}{\small H}_{0}$ \\ \cline{9-11}
&  &  &  &  &  &    &  & ${\small z=1-3}$ &  &   ${\small z=1090}$ &  &  
\\ \hline
$\Lambda ${\small CDM} &  &  & {\small 0.671} &  &  & {\small 0.333}   &  &
\multicolumn{1}{l}{\small 0.226--0.035} &  &   \multicolumn{1}{c}{\small %
0.000} &  &   {\small 0.964} \\
{\small QUINT} &  &  & {\small 0.638} &  &  & {\small 0.417}   &  &
\multicolumn{1}{l}{\small 0.307--0.077} &  &   \multicolumn{1}{c}{\small %
0.000} &  &   {\small 0.931} \\
{\small Phantom} &  &  & {\small 0.650} &  &  & {\small 0.278}   &  &
\multicolumn{1}{l}{\small 0.161--0.016} &  &   \multicolumn{1}{c}{\small %
0.000} &  &   {\small 0.991} \\
{\small UIS} &  &  & {\small 0.441} &  &  & {\small 0.392}   &  &
\multicolumn{1}{l}{\small 0.276--0.208} &  &   \multicolumn{1}{c}{\small %
0.011} &  &   {\small 0.934} \\
{\small FSLL1} &  &  & {\small 0.699} &  &  & {\small 0.326}   &  &
\multicolumn{1}{l}{\small 0.227--0.034} &  &   \multicolumn{1}{c}{\small %
0.000} &  &   {\small 0.976} \\
{\small FSLL2} &  &  & {\small 0.698} &  &  & {\small 0.325}   &  &
\multicolumn{1}{l}{\small 0.226--0.033} &  &   \multicolumn{1}{c}{\small %
0.000} &  &   {\small 0.977} \\
{\small BA} &  &  & {\small 0.810} &  &  & {\small 0.348}   &  &
\multicolumn{1}{l}{\small 0.286--0.051} &  &   \multicolumn{1}{c}{\small %
0.000} &  &   {\small 0.998} \\
{\small MZ} &  &  & {\small 0.676} &  &  & {\small 0.349}   &  &
\multicolumn{1}{l}{\small 0.246--0.054} &  &   \multicolumn{1}{c}{\small %
0.000} &  &   {\small 0.983} \\
{\small PSY1} &  &  & {\small 0.665} &  &  & {\small 0.325}   &  &
\multicolumn{1}{l}{\small 0.214--0.028} &  &   \multicolumn{1}{c}{\small %
0.000} &  &   {\small 0.961} \\
{\small PSY2} &  &  & {\small 0.644} &  &  & {\small 0.338}   &  &
\multicolumn{1}{l}{\small 0.222--0.035} &  &   \multicolumn{1}{c}{\small %
0.000} &  &   {\small 0.952} \\ \hline\hline
\end{tabular}%
\end{table}

Due to tension in present values of $H_{0}$ \cite{ries22,hu22}, the age of 
the universe $tH_{0}$ is presented in the last column of
Table 1. As $tH_{0}$ is an integrated observable, the dynamical evolution of
dark energy vis-a-vis the signature and the functional form of $w(z)$ leaves
an indelible imprint on the age of the universe. The age $tH_{0}$ is
close to $1$ with BA parametrizations. In quintessence model, the value of $%
tH_{0}$ is maximally off by 7\%.

In Table 2, we display the future evolution of dark energy parameter for $%
a=2,4$ and $1000$. It is noticed that in comparison to $\Lambda $CDM, the
variation in $\Omega _{X}$ due to all the considered parametrizations 
at $a=2$ and $4$
lies within 4\%. The dark energy parameter $\Omega _{X}=1$ at $a=1000$ in
all the parametric models. In fact, $\Omega _{X}$ is equal to unity as early
as $a=10$. However, a large variation is observed in $t_{1000}H_{0}$. The
smallest and largest values of $t_{1000}H_{0}$ are 0.901 and 27.556 with UIS
and quintessence parametrizations, respectively.

\begin{table}[htbp] 
\caption{ Evolution of dark energy parameter $\Omega _{X}$ in the future
epoch.}
\begin{tabular}{lllllllllllllllll}
\hline\hline
{\small Model} &  &  &  &  &  &  &  & ${\small \Omega }_{X}$ &  &  &  &  &
&  &  & ${\small t}_{1000}{\small H}_{0}$ \\ \cline{5-12}
&  &  &  &  & ${\small a=2}$ &  &  & ${\small a=4}$ &  &  &
\multicolumn{1}{c}{${\small a=1000}$} &  &  &  &  &  \\ \hline
$\Lambda ${\small CDM} &  &  &  &  & {\small 0.949} &  &  & {\small 0.993} &
&  & \multicolumn{1}{c}{\small 1.000} &  &  &  &  & {\small 8.182} \\
{\small QUINT} &  &  &  &  & {\small 0.925} &  &  & {\small 0.985} &  &  &
\multicolumn{1}{c}{\small 1.000} &  &  &  &  & {\small 27.56} \\
{\small Phantom} &  &  &  &  & {\small 0.966} &  &  & {\small 0.997} &  &  &
\multicolumn{1}{c}{\small 1.000} &  &  &  &  & {\small 3.426} \\
{\small UIS} &  &  &  &  & {\small 0.987} &  &  & {\small 1.000} &  &  &
\multicolumn{1}{c}{\small 1.000} &  &  &  &  & {\small 0.901} \\
{\small FSLL1} &  &  &  &  & {\small 0.952} &  &  & {\small 0.994} &  &  &
\multicolumn{1}{c}{\small 1.000} &  &  &  &  & {\small 7.799} \\
{\small FSLL2} &  &  &  &  & {\small 0.954} &  &  & {\small 0.994} &  &  &
\multicolumn{1}{c}{\small 1.000} &  &  &  &  & {\small 7.140} \\
{\small BA} &  &  &  &  & {\small 0.954} &  &  & {\small 0.993} &  &  &
\multicolumn{1}{c}{\small 1.000} &  &  &  &  & {\small 10.67} \\
{\small MZ} &  &  &  &  & {\small 0.964} &  &  & {\small 0.997} &  &  &
\multicolumn{1}{c}{\small 1.000} &  &  &  &  & {\small 2.930} \\
{\small PSY1} &  &  &  &  & {\small 0.950} &  &  & {\small 0.994} &  &  &
\multicolumn{1}{c}{\small 1.000} &  &  &  &  & {\small 4.312} \\
{\small PSY2} &  &  &  &  & {\small 0.944} &  &  & {\small 0.992} &  &  &
\multicolumn{1}{c}{\small 1.000} &  &  &  &  & {\small 9.227} \\ \hline\hline
\end{tabular}
\end{table}

\begin{figure*}[htbp]
\begin {minipage}[t]{7cm}
\centering
\includegraphics[scale=0.32] {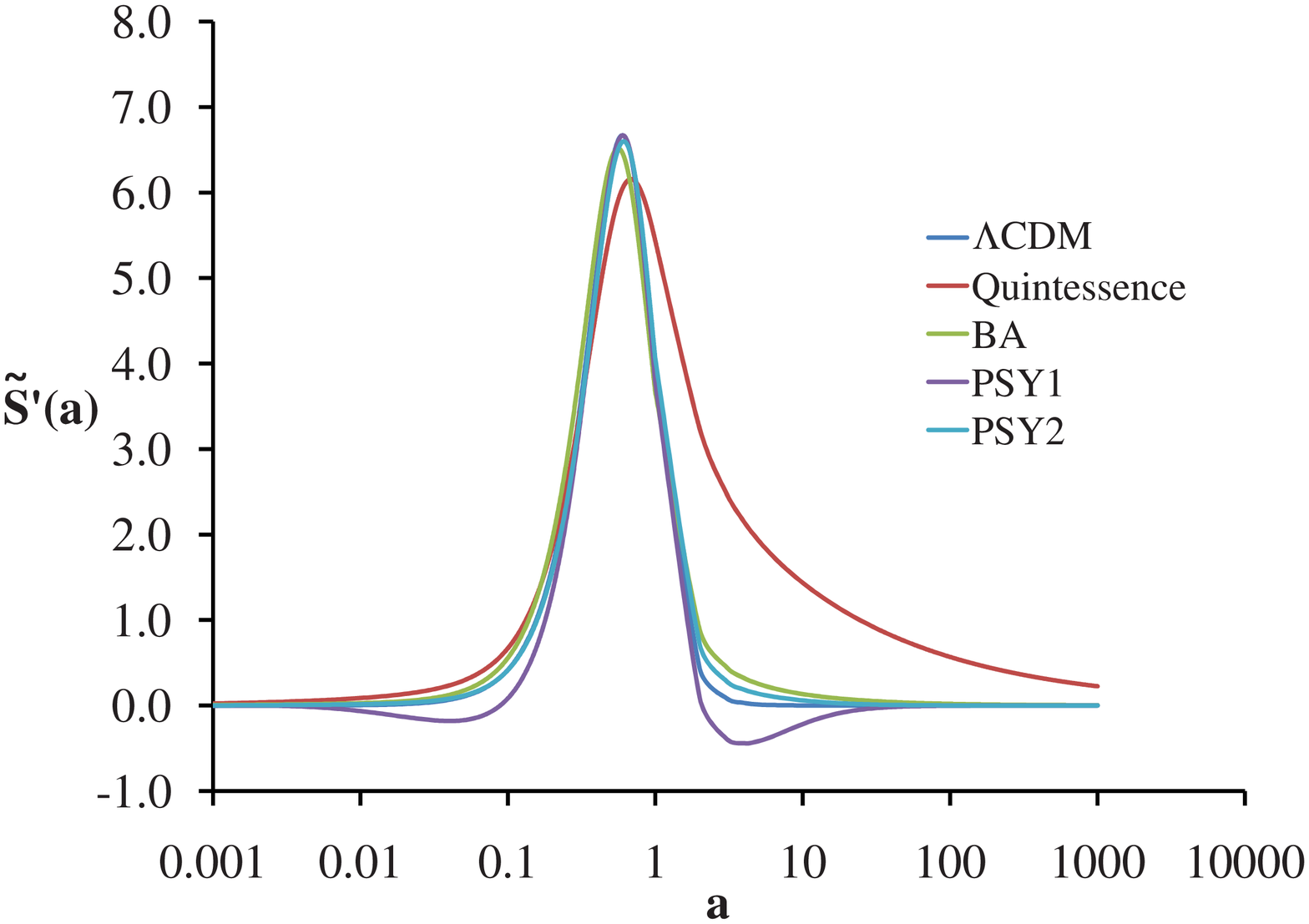}
\caption{ Evolution of $\widetilde{S}^{\prime}(a)$  with $\Lambda $CDM, Quintessence,
BA, PSY1 and PSY2 parametrizations.}
\label{fig1}
\end{minipage}
\hspace{1cm}
\begin{minipage}[t]{7cm}
\centering
\includegraphics[scale=0.32] {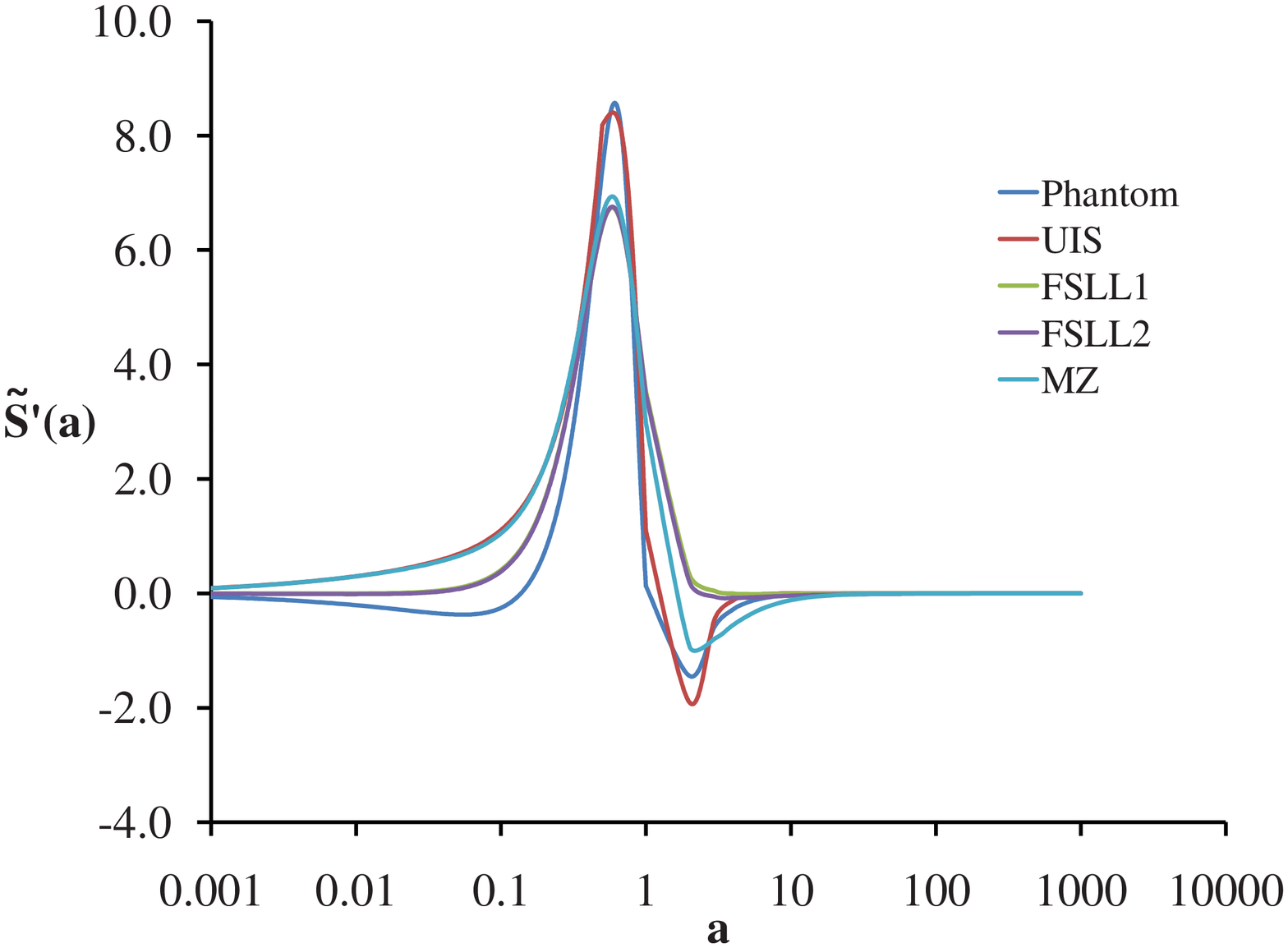}
\caption{Evolution of $\widetilde{S}^{\prime}(a)$ with Phantom, UIS, FSLL1,
FSLL2 and MZ parametrizations.}
\label{fig2}
\end{minipage}
\end{figure*}

Thermodynamic viability of GSL is given by $S^{\prime }(a)\geqslant 0$ and
in the far future $S^{\prime \prime }(a)\leq 0$. Hence, it is required that $%
S^{\prime }(a)\geqslant 0$ through out the dynamical evolution of the
universe i.e. $a=0-\infty $. As $\Omega _{X}\rightarrow 0$ for $a<0.0009$ (BBN)
and $\Omega_{X}=1$ for $a=10$, the behavior of $S^{\prime }(a)$ is investigated
within range $a=0.001$--$1000$ corresponding to $z_{CMB}=1090$ and redshift
at far future $z=-0.999$ as asymptotic limit $a\rightarrow \infty $. In
general, the contribution of $\widetilde{S}_{M}^{\prime }(a)$ and $%
\widetilde{S}_{R}^{\prime }(a)$ to total $\widetilde{S}^{\prime }(a)$ is
negligible. The evolution of $\widetilde{S}^{\prime }(a)$ is governed by the
relative contribution of $\widetilde{S}_{H}^{\prime }(a)$ and $\widetilde{S}%
_{X}^{\prime }(a)$. In $\Lambda $CDM model, the $\widetilde{S}_{X}^{\prime
}(a)=0$ due to the $(1+w_{X})$ term in Eq. (\ref{ssrx}) albeit the dark
energy parameter $\Omega _{X}$ contributes to the horizon entropy $%
S_{H}^{{}}(a)$. In the vicinity of $a=0.001$, $\widetilde{S}_{H}^{\prime
}(a)\sim 0$ as $\Omega _{X}\sim 0$. As $a\rightarrow 1$, 
there is an increase in $\widetilde{S}_{H}^{\prime }(a)$.
Further, $\widetilde{S}_{H}^{\prime }(a)\rightarrow 0$ as $\Omega
_{X}\sim 1$ around $a=10$. The second condition $\left. S^{\prime \prime
}(a)\right\vert _{a=1000}\leq 0$ is also satisfied as $\widetilde{S}%
_{H}^{\prime }(a)$ decreases with the increase of $a\rightarrow 1000$.

With other parametrizations, relative contributions of horizon and dark
energy entropy decide the evolution of total entropy. The signs of $%
\widetilde{S}_{H}^{\prime }(a)$ and $\widetilde{S}_{X}^{\prime }(a)$ are
decided by the terms ($1+w_{X}(a)\Omega _{X}(a)$) and ($1+3w_{X}(a)\Omega
_{X}(a)$) in Eq.(\ref{ssh}) and Eq.(\ref{ssrx}), respectively. Further, the
magnitude of $\widetilde{S}_{X}^{\prime }(a)$ and hence $\widetilde{S}%
^{\prime }(a)$ is modulated by the parameter $\tau _{X}$ as shown in Ref. 
\cite{card17}. Presently, $\tau _{X}=1$ is used. In Figs. 1 and 2, variation
of $\widetilde{S}^{\prime }(a)$ vs. $a$ due to different parametrizations is
plotted within range $a=0.001$--$1000$. In all cases, the peak in
variation of $\widetilde{S}^{\prime }(a)$ is observed to be at the transition epoch
corresponding $z_{T}$. This is expected as the sign  of $\Omega _{X}(a)$ changes 
at $a_{T}$. It is also observed that models
presented in Fig. 1, namely $\Lambda $CDM, quintessence, BA, PSY1 and PSY2
have $\widetilde{S}^{\prime \prime }(a)<0$ and $\widetilde{S}^{\prime }(a)$
due to phantom, UIS, FSSL1, FSSL2 and MZ parametrizations displayed in Fig.
2 have $\widetilde{S}^{\prime \prime }(a)>0$. However, thermodynamic
viability of parametric models cannot be decided due to the unknown value of
arbitrary parameter $\tau _{X}$.



\section{Conclusions}

A comparative study of a set of parametric dark energy models has been
performed by calculating the dynamical evolution of dark energy parameter $%
\Omega _{X}$ in the past and future epochs. In comparison to $\Lambda $CDM
model, the variation in $\Omega _{X}$ due to different parametrization is
about 5\% in the expansion parameter range $a=10^{-14}-1000$. The maximum
variation in the age of the universe $t_{0}H$ is about 7\%. However, $%
t_{1000}H$ (time till $a=1000$) varies by a factor of 30.

Thermodynamic viability of dark energy models has been investigated by
calculating the two requirements of GSL of thermodynamics i.e. $S^{\prime
}(a)$ and $S^{\prime \prime }(a)$. It has been observed that with reasonable
assumptions, GSL of thermodynamics is satisfied by $\Lambda $CDM,
quintessence, BA, PSY1 and PSY2 parametrizations. Parametrizations having
phantom behavior are not suitable from thermodynamic perspective.
Thermodynamic viability of parametric models cannot be conclusively
ascertained to the unknown value of $\tau _{X}$.


The author gratefully acknowledges fruitful discussions with Prof. T. R. Seshadri, 
{\small 
Department of Physics and Astrophysics, University of Delhi, }India. Additionally, the author 
expresses her gratitude to Dr. Ramesh Chandra, Department of Physics, Babasaheb Bhimrao
Ambedkar University, Lucknow, for his guidance throughout the completion of the present work. 

\end{document}